\documentclass[11pt,twoside]{stm}

\renewcommand{\thesection}{\Roman{section}}

\newcommand{\gp}{g^{\prime}}

\newcommand{\diag}{\mbox{diag}}

\newcommand{\rhoslash}{\not{\!\! \rho}}
\newcommand{\Bslash}{\not{\!\! B}}
\newcommand{\Wslash}{\not{\!\! W}}
\newcommand{\eqref}[1]{(\ref{#1})}

\begin{document}
\markboth{\sc M.~Gintner, I.~Melo, B.~Trpi\v{s}ov\'{a}}
{\sc Probing the strong electroweak symmetry breaking}

\STM

\title{Probing the strong electroweak symmetry breaking
in a model with a vector resonance}

\authors{M.~Gintner$^{1,2}$, I.~Melo$^1$, B.~Trpi\v{s}ov\'{a}$^1$}

\address{$^1$ Physics Department, University of \v{Z}ilina, \v{Z}ilina\\
$^2$ Science and Research Institute, Matej Bel University, Bansk\'{a} Bystrica}
\bigskip

\begin{abstract}
We systematically study the possibility to probe the physics behind the electroweak
symmetry breaking at the LHC assuming new strong interactions being responsible
for the effect. The new physics is described by the Higgs-less effective Lagrangian with a vector
resonance in the particle spectrum, in addition to the Standard Model fields. We
analyze signals of the resonance in the mixing-induced LHC processes
$pp\rightarrow abX$, $ab=t\bar{t},b\bar{b},t\bar{b},W^+Z,W^+W^-$. At
this level of our calculations we do not consider further decays of the quarks and
of the gauge bosons.
\end{abstract}

\section*{Introduction}

Despite the great success of the Standard model of electroweak
interactions (SM) one essential component of the theory remains
a puzzle: it is the actual mechanism behind the electroweak symmetry breaking (ESB).
The Higgs complex doublet scalar field of a non-zero vacuum expectation
value serves as a benchmark hypothesis for the mechanism. It is not the only
plausible hypothesis though. A direct
consequence of the Higgs hypothesis is the presence of the Higgs boson
in the particle spectrum of the SM, not observed to these days.

There is a host of candidates for the suitable extensions of the SM.
They range from supersymmetric theories with multiple elementary Higgs bosons
in their spectra to the theories of new strong interactions, like in Technicolor,
which might form bound states of new elementary particles in analogy with QCD.
These bound states might appear in the particle spectrum as new resonances.

In more recent theories like the Little Higgs
models and the Gauge-Higgs unification models, their dual-description relation
to the Heavy Composite Higgs and the No Higgs strongly-interacting models has
been demonstrated (see \cite{Cheng2007} and references therein).
Most of these new models introduce
new quarks and new vector particles at about 1~TeV.

Facing this plethora of alternative theories it is highly desirable
to describe their low-energy phenomenology in a unified way.
Thus, it is very useful to exploit the formalism of effective Lagrangians.
We use the so-called {\it hidden local symmetry} (HLS) approach \cite{Bando1988}
to introduce the new vector resonance.
The HLS formalism along with the AdS/CFT correspondence plus deconstruction
is also behind the dual-description relation of the recent models mentioned above
\cite{Cheng2007}.

In our model, the new $SU(2)_V$ vector bosons mix with the electroweak gauge bosons.
While the mixing complicates theoretical
analysis of the model it might provide some advantages in experimental
searches. It is natural to assume that apart from
the electroweak gauge bosons the new vector triplet couples to the third
generation of quarks only.
Therefore it would seem natural to search for the signs of the new resonances
in the processes where besides the electroweak bosons the $t$ and $b$ quarks are involved.
At the LHC this is often a difficult case due to the large backgrounds to the processes
with $b$ and $t$ in the final state and/or due to the negligible $b$ and $t$ luminosities in
$pp$ collisions. However, the mixing of the vector resonances
with the electroweak gauge bosons generates interactions
of the new resonances to the fermions of lighter generations.
Although these interactions are suppressed by the mixing factors,
the processes enabled by them have the advantage of higher parton luminosities and
more favorable final state topologies, on the other hand.

In the next section we briefly introduce our effective
Lagrangian. In the following section
we present results of our analysis. In Conclusions we summarize our findings and
outline the prospective steps of this investigation.

\section*{Lagrangian}
\label{sec:Lagrangian}

Our Lagrangian is based on the
non-linear Callan-Coleman-Wess-Zumino Lagrangian \cite{CWZ1969,CCWZ1969}.
This approach was applied in formulating the so-called
BESS (Breaking Electroweak Symmetry Strongly) model (e.g. \cite{BESS1991}
and references therein) which
pioneered the use of the HLS approach to the effective description of
new resonances formed by new strong interactions related to the ESB.

The BESS model considered in the literature
possesses the $\rho$-to-fermion coupling inter-generation
universality \cite{BESS1991} which leads to stringent limits
on these couplings from the existing measurements of the SM parameters.
In an attempt to reflect the speculations about a special role of the top
quark (or the third quark generation) in the mechanism of ESB
we modify the fermion sector by considering no direct interactions
of $\rho$ to the fermions other than $t$ and $b$. In addition,
the $SU(2)_L$ symmetry does not allow us to disentangle the $\rho$-to-$t_L$
coupling from the $\rho$-to-$b_L$ one. However, it can be done in the case
of right fermions. We do it and to simplify the numerical analysis of the model
we turn off  completely the direct interaction of $\rho$ to $b_R$.
The full form of our Lagrangian can be found in \cite{HK2008}.
Here, we show only the part responsible for interactions of $t$ and $b$ quarks
to the new vector resonances as well as non-SM interactions of $t$ and $b$ to the SM gauge bosons
\begin{eqnarray}
  L_\rho^{(t,b)_L} &=& \frac{b_1}{1+b_1}g \;\bar{\psi}_L\!\Wslash_{\! a}\tau_a\psi_L
  + \frac{b_1}{1+b_1} g_V \;\bar{\psi}_L\!\rhoslash_{\! a}\tau_a\psi_L
  \label{LrhoL}\\
  L_\rho^{(t,b)_R} &=& \frac{b_2}{1+b_2}\gp (\bar{\psi}_R P_0)\Bslash\tau_3(P_0\psi_R)
  + \frac{b_2}{1+b_2} g_V (\bar{\psi}_R P_0)\rhoslash_3\tau_3(P_0\psi_R)
  \label{LrhoR}
\end{eqnarray}
where $\psi=(t,b)^T$, and $P_0=\diag(1,0)$. As a consequence of turning off
the $\rho$-to-$b_R$ coupling neither $t_R$ directly couples to the charged
$\rho$-resonance, as can be seen in \eqref{LrhoR}.

The measurement of $Zb\bar{b}$
vertex constrains the $\rho$-to-$t_L$ coupling to relatively small values.
However, when the disentangling is applied the
measurement does not limit the $\rho$-to-$t_R$ interaction.
For the low-energy limits on the model parameters, see \cite{GMT2006}.

\section*{Two-particle final state processes: analysis and results}
\label{sec:TwoParticleFSP}

\begin{figure}[h]
\includegraphics[scale=0.27]{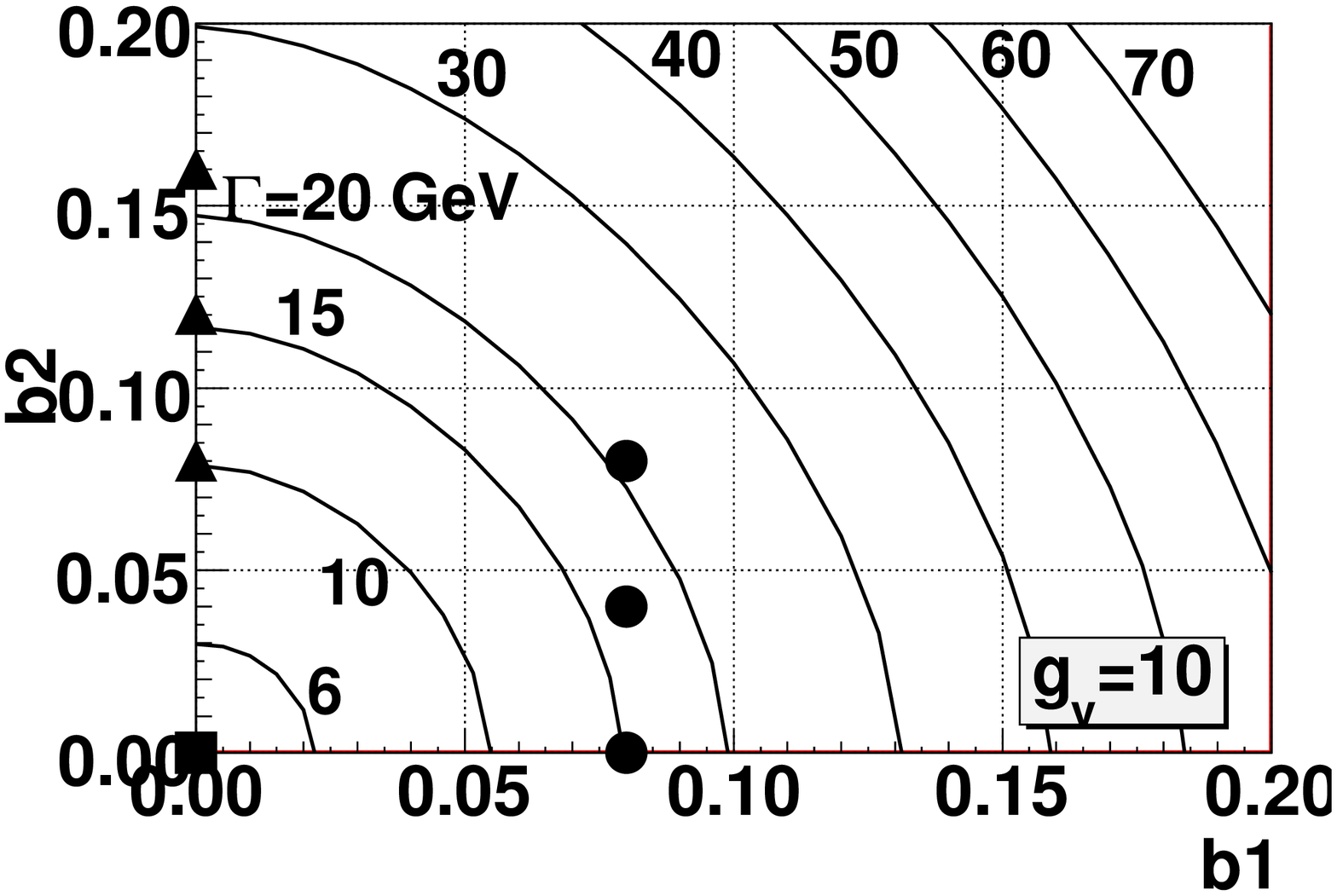}
\includegraphics[scale=0.27]{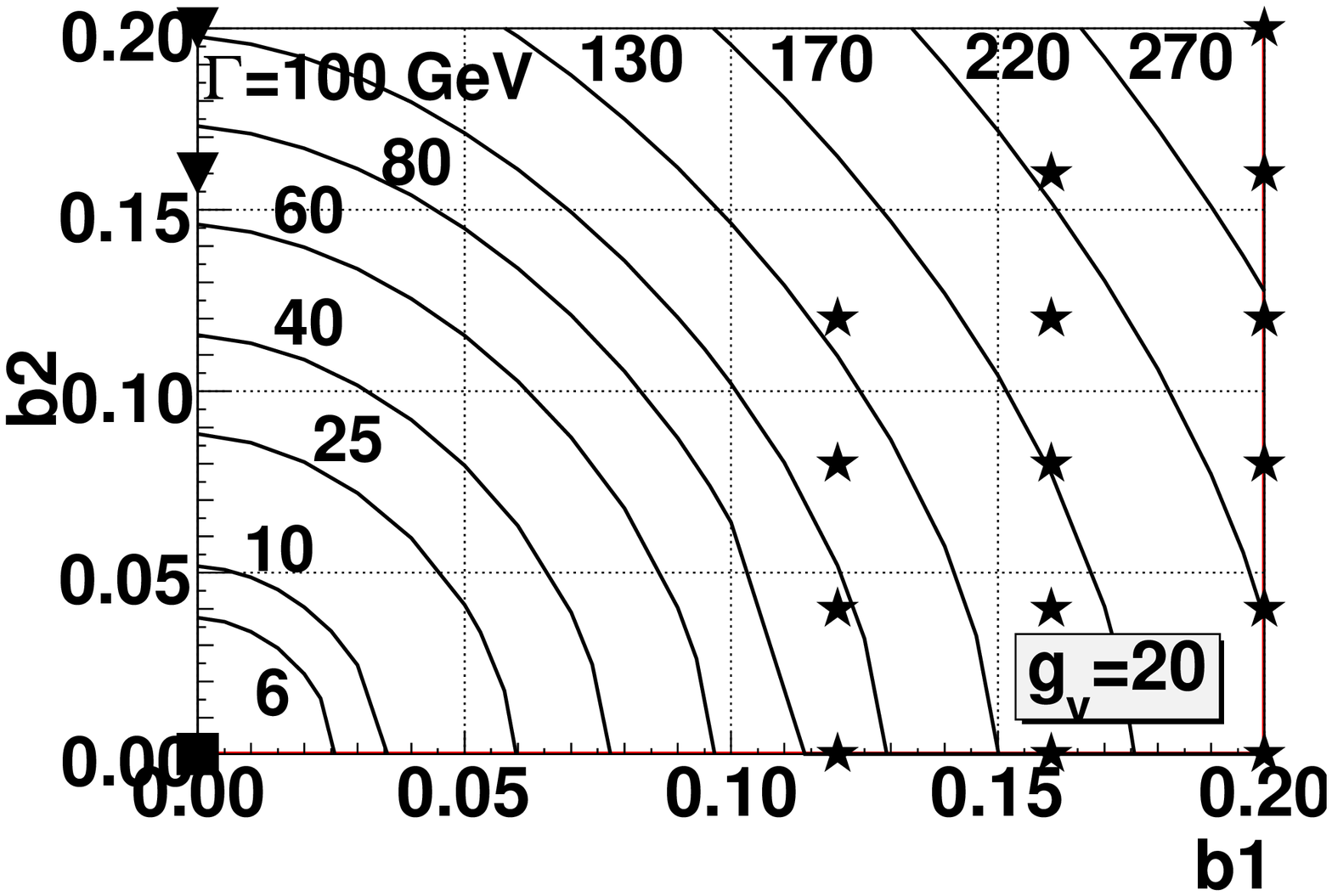}
\includegraphics[scale=0.27]{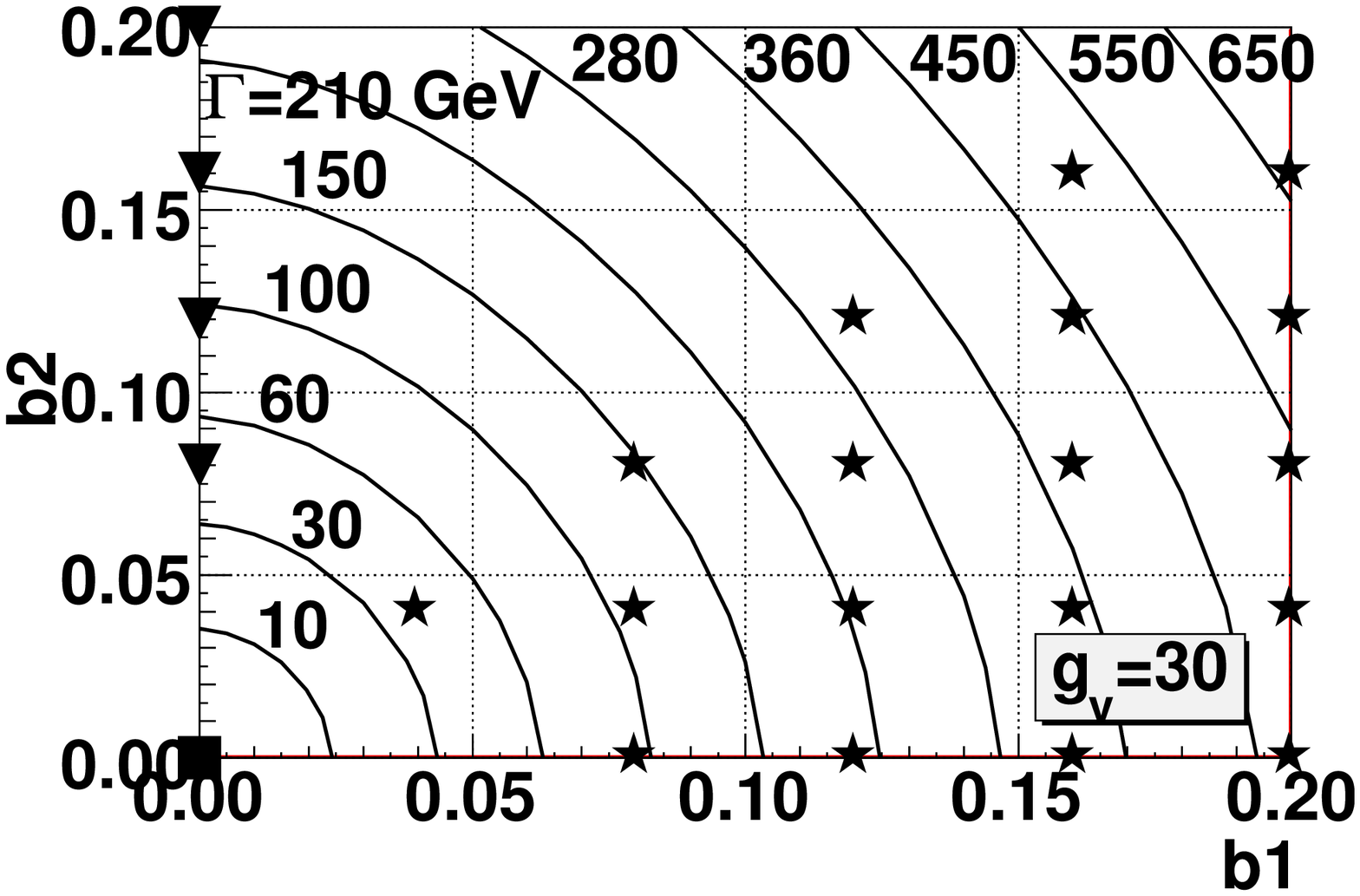}
\caption{The total widths of $\rho^0$ depicted in
the $g_V-b_1-b_2$ parametric space. The markers distinguish areas
of a clear domination of some channels; circles: $t\bar{t}\sim b\bar{b}\sim W^+W^-$,
squares: $W^+W^-\gg t\bar{t},b\bar{b}$, down-pointing triangles: $t\bar{t}\gg b\bar{b},W^+W^-$,
stars: $t\bar{t},b\bar{b}\gg W^+W^-$, up-pointing triangles: $t\bar{t},W^+W^-\gg b\bar{b}$.}
\label{fig:ParamSpace}
\end{figure}

We test the model when the mass of the neutral $\rho$-resonance is
$M_{\rho^0}=1$~TeV. Depending on the model parameters $\rho^0$ decays
predominantly through one or more of the following three channels:
$t\bar{t}$, $b\bar{b}$, $W^+W^-$. When its mass is fixed the total width
of $\rho^0$, as well as its branching ratios, depend on $g_V$, $b_1$,
and $b_2$. The situation in the $g_V-b_1-b_2$ parametric
space is depicted in Figure~\ref{fig:ParamSpace}.
Note that due to the turning off of the $\rho$-to-$b_R$ coupling
there is no such a choice of parameters when the decay of $\rho^0$ either solely
to $b\bar{b}$ or to both, $b\bar{b}$ and $W^+W^-$, dominates.

The parameters $M_{\rho^0}$, $g_V$, $b_1$, and $b_2$ also determine
the mass of $\rho^\pm$ and its decay width and branching ratios.
The $\rho^\pm$-resonance
dominantly decays to $t\bar{b}/\bar{t}b$ and/or $W^\pm Z$.
In Table \ref{tab:PSP} we show the points of the $g_V-b_1-b_2$ parametric space
we test in our calculations.
\begin{table}
  \begin{tabular}{|c|c|c|c|r|c|c|c|c|r|c|c|}
    \hline
    P & $g_V$ & $b_1$ & $b_2$ & \multicolumn{1}{c|}{$\Gamma_{\rho^0}\;$} &
    \multicolumn{3}{c|}{BR($\rho^0$)} & $M_{\rho^\pm}\;$
    & \multicolumn{1}{c|}{$\Gamma_{\rho^\pm}\;$}
    & \multicolumn{2}{c|}{BR($\rho^\pm$)}\\
    \cline{6-8}\cline{11-12}
    & & & & (GeV) & $W^+W^-$ & $t\bar{t}$ & $b\bar{b}$ & \multicolumn{1}{c|}{(GeV)}
    & \multicolumn{1}{c|}{(GeV)} & $t\bar{b}/\bar{t}b$ & $W^\pm Z$ \\
    \hline\hline
    1 &  10 & 0.08 & 0.04 & 16.899 & 31\% & 38\% & 31\% & 999.84 & 15.281 & 64\% & 36\% \\
    2 &  10 & 0.12 & 0.04 & 28.256 & 19\% & 42\% & 39\% & 999.84 & 26.433 & 79\% & 21\% \\
    \hline
    3 &  10 & 0 & 0 & 5.334 & 99\% & 0.12\% & 0.08\% & 999.84 & 5.443 & 0.2\% & 98\% \\
    \hline
    4 &  20 & 0 & 0.12 & 42.788 & 3\% & 97\% & 0.0025\% & 999.96 & 1.358 & 0.2\% & 98\% \\
    \hline
    5 &  20 & 0.08 & 0 & 42.471 & 3\% & 46\% & 51\% & 999.96 & 42.509 & 97\% & 3\% \\
    6 &  35 & 0.04 & 0 & 34.580 & 1\% & 47\% & 52\% & 999.99 & 34.594 & 99\% & 1\% \\
    \hline
    7 &  10 & 0 & 0.08 & 10.169 & 52\% & 48\% & 0.042\% & 999.84 & 5.443 & 0.18\% & 98\% \\
    \hline
  \end{tabular}
\caption{Properties of the $\rho$-triplet in the selected points of the
$g_V-b_1-b_2$ parametric space when $M_{\rho^0}=1$~TeV.}
\label{tab:PSP}
\end{table}

We begin with the analysis of two-particle processes not only because
they are the simplest ones to calculate but also to demonstrate that
the visible $\rho$-resonance peaks can appear even in the presence of
the mixing-suppressed couplings. We implemented our Lagrangian \cite{HK2008} into
the CompHEP software \cite{CompHEP1999,CompHEP2004}
as one of its models. Numerical analysis was performed
at the selected points of the parametric space.
All calculations have been performed at tree level.

There have been five two-particle processes
analyzed. They are listed in Table \ref{tab:processes2}.
The cuts have been introduced to maintain numerical stability of the
calculations or to simulate the blind angle of the detector. No attempt
to improve the signal-to-background ratio using cuts has been made.
\begin{table}[h]
  \begin{tabular}{|c|c|c|p{1.2cm}|c|c|}
    \hline
    process & subprocess & P & $\rho$-peak visible & $\sigma\;$(pb) & cuts \\
    \hline\hline
    $pp\rightarrow t\bar{t}X$ & $gg\rightarrow t\bar{t}$ & 1--7 & no & 726 & no \\
    \cline{2-5}
     & $b\bar{b}\rightarrow t\bar{t}$& 6  & yes & 1.69 &  \\
    \hline
    $pp\rightarrow b\bar{b}X$ & $gg\rightarrow b\bar{b}$ & 1--7 & no & 1120
    & $m_{b\bar{b}}\geq 350$~GeV \\
    \cline{2-5}
     & $b\bar{b}\rightarrow b\bar{b}$& 5 & yes & 22.5
    & $-0.95\leq c_{13},c_{14}\leq 0.95$ \\
    \hline
     & $u\bar{d}\rightarrow t\bar{b}$ & 2 & yes & 1.96
    &  \\
    \cline{2-2}\cline{4-5}
    & $c\bar{s}\rightarrow t\bar{b}$ & & yes & 0.29 & \\
    \cline{2-5}
    $pp\rightarrow t\bar{b}X$ & $u\bar{d}\rightarrow t\bar{b}$ & 5 & yes & 1.99 & $-0.99\leq c_{13},c_{14}\leq 0.99$\\
    \cline{2-2}\cline{4-5}
    & $c\bar{s}\rightarrow t\bar{b}$ & & yes & 0.30 & \\
    \cline{2-5}
    & $u\bar{d}\rightarrow t\bar{b}$ & SM & N/A & 1.85 & \\
    \cline{2-2}\cline{4-5}
    & $c\bar{s}\rightarrow t\bar{b}$ & & N/A & 0.28 & \\
    \hline
     & $u\bar{u}\rightarrow W^+W^-$ & 2 & no & 14.02 & \\
    \cline{2-2}\cline{4-5}
    & $b\bar{b}\rightarrow W^+W^-$ & & yes & 0.77 & \\
    \cline{2-5}
    $pp\rightarrow W^+W^-X$ & $u\bar{u}\rightarrow W^+W^-$ & 3 & yes & 14.02 & $-0.99\leq c_{13},c_{14}\leq 0.99$\\
    \cline{2-2}\cline{4-5}
    & $b\bar{b}\rightarrow W^+W^-$ & & yes & 0.74 & \\
    \cline{2-5}
    & $u\bar{u}\rightarrow W^+W^-$ & SM & N/A & 13.13 & \\
    \cline{2-2}\cline{4-5}
    & $b\bar{b}\rightarrow W^+W^-$ & & N/A & 0.70 & \\
    \hline
    &  $u\bar{d}\rightarrow W^+Z$ & 1 & yes & 5.27 & \\
    \cline{2-2}\cline{4-5}
    & $c\bar{s}\rightarrow W^+Z$ &  & yes & 0.91 & \\
    \cline{2-5}
    $pp\rightarrow W^+ZX$ & $u\bar{d}\rightarrow W^+Z$ & 3 & yes & 5.28 & $-0.99\leq c_{13},c_{14}\leq 0.99$\\
    \cline{2-2}\cline{4-5}
    & $c\bar{s}\rightarrow W^+Z$ & & yes & 0.91 & \\
    \cline{2-5}
    & $u\bar{d}\rightarrow W^+Z^-$ & SM & N/A & 4.60 & \\
    \cline{2-2}\cline{4-5}
    & $c\bar{s}\rightarrow W^+Z^-$ & & N/A & 0.79 & \\
    \hline
  \end{tabular}
\caption{Properties of the analyzed two-particle final state processes at $\sqrt{s}=14$~TeV.
Only the most important and interesting subprocesses and points P of the parametric space
are displayed. The cross section values correspond to the
$q\bar{q}'+\bar{q}'q$ initial states of the subprocesses. The SM results are calculated
assuming $M_{Higgs}=115$~GeV; $c_{13}$ and $c_{14}$ denote cosines of the scattering angles of
the first and the second final state particles, respectively.}
\label{tab:processes2}
\end{table}

We would like to know if the LHC signal by our model can be distinguished
from the LHC signal of the SM with the Higgs boson of the mass
$M_{Higgs}=115$~GeV. For that purpose we calculate the statistical
significance of the model signal with respect to the SM
\[ R=\frac{N_{P}-N_{SM}}{\sqrt{N_{SM}}} \]
where $N_{P}$ and $N_{SM}$ are the numbers of the events of our model and
the SM, respectively. It is customary to consider as statistically
significant such a deviation for which $R>5$.
We calculate $R$'s based on the total cross sections
as well as on the cross sections in the vicinity of the $\rho$ peak,
$0.7\;\mbox{TeV}\leq m_{34}\leq 1.1\;\mbox{TeV}$, where $m_{34}$ is the
final state pair of particles invariant mass.
No reducible backgrounds and no improving
cuts are considered at this stage.
Since $pp\rightarrow t\bar{t}X$ and
$pp\rightarrow b\bar{b}X$ are plagued by gluon-gluon backgrounds
we focus on the remaining three processes in our calculations.
The obtained results are
shown in Table \ref{tab:BeforeDecayR}.
\begin{table}
\begin{center}
  \begin{tabular}{|c|c|c|c|c|c|}
    \hline
    process & P & cut & $\sigma\;$(pb) & $R_0$ & $R$ \\
    &&&&& $(100\;\mbox{fb}^{-1})$ \\
    \hline\hline
     & SM & no & 5.84 & 0 & 0 \\
    \cline{2-2}\cline{4-6}
    $pp\rightarrow t\bar{b}X + c.c$ & 2 & & 6.17 & 0.136 & 43.04 \\
    \cline{2-6}
     & SM & $0.7\;\mbox{TeV}\leq m_{tb}\leq 1.1\;\mbox{TeV}$ & 0.14 & 0 & 0 \\
    \cline{2-2}\cline{4-6}
     & 2 & & 0.20 & 0.163 & 51.47 \\
    \hline
     & SM & no & 14.77 & 0 & 0 \\
    \cline{2-2}\cline{4-6}
    $pp\rightarrow W^+ZX + c.c$ & 3 & & 16.96 & 0.570 & 180.37 \\
    \cline{2-6}
     & SM & $0.7\;\mbox{TeV}\leq m_{WZ}\leq 1.1\;\mbox{TeV}$ & 0.20 & 0 & 0 \\
    \cline{2-2}\cline{4-6}
     & 3 & & 0.29 & 0.188 & 59.30 \\
    \hline
     & SM & no & 29.86 & 0 & 0 \\
    \cline{2-2}\cline{4-6}
    $pp\rightarrow W^+W^-X$ & 3 & & 31.86 & 0.366 & 115.74 \\
    \cline{2-6}
     & SM & $0.7\;\mbox{TeV}\leq m_{WW}\leq 1.1\;\mbox{TeV}$ & 0.37 & 0 & 0 \\
    \cline{2-2}\cline{4-6}
     & 3 & & 0.42 & 0.097 & 30.75 \\
    \hline
  \end{tabular}
\end{center}
\caption{Cross sections and statistical significance $R$ of the model signals
with respect to the SM for the studied processes when the integrated luminosity
${\cal L}=100\;\mbox{fb}^{-1}$. $R_0 = (\sigma_P-\sigma_{SM})/\sqrt{\sigma_{SM}}$.}
\label{tab:BeforeDecayR}
\end{table}

Since none of the final state particles of the studied processes is stable,
it is more relevant to consider the statistical significance when the decays
are taken into account. In addition, also the question of detection efficiency
of a given final state is important in evaluating more realistic values of $R$.
In our calculations we consider the following reduction factors for the
observed number of events: the efficiency of the $b$-jet detecting $\epsilon_b=0.5$;
the branching ratio $b_{jj}^W=\;$BR$(W\rightarrow jj)=0.64$, where $j$ is a
light-quark jet; $b_{\ell\nu}^W=\;$BR$(W\rightarrow \ell\nu_\ell)=0.11$,
where $\ell$ is one of the charged SM leptons;
$b_{\ell\ell}^Z=\;$BR$(Z\rightarrow \ell^-\ell^+)=0.034$;
$b_{jj}^Z=\;$BR$(Z\rightarrow jj)=0.538$;
$b_{bb}^Z=\;$BR$(Z\rightarrow b\bar{b})=0.153$.
The obtained statistical significance for the processes after the final state
decay is shown in Table \ref{tab:AfterDecay}.
\begin{table}
\begin{center}
  \begin{tabular}{|c|c|c|c|c|c|}
    \hline
    final state & reduction & P & cut  & events & $R$ \\
    & factor $r$ &&& $(100\;\mbox{fb}^{-1})$ & $(100\;\mbox{fb}^{-1})$ \\
    \hline\hline
    \multicolumn{6}{|c|}{$pp\rightarrow t\bar{b}X + c.c.$}\\
    \hline
    $\ell^+\nu_\ell b\bar{b}+c.c$ & $\epsilon_b^2 b_{\ell\ell}^W$ & 2 & no & $1.70\times 10^4$ & 7.14 \\
    \cline{4-6}
    & 0.0275 && yes & $5.40\times 10^2$ & 8.53 \\
    \hline
    $jj b\bar{b}+c.c$ & $\epsilon_b^2 b_{jj}^W$ & 2 & no & $9.86\times 10^4$ & 17.22 \\
    \cline{4-6}
    & 0.16 && yes & $3.14\times 10^3$ & 20.59 \\
    \hline\hline
    \multicolumn{6}{|c|}{$pp\rightarrow W^+ZX + c.c.$}\\
    \hline
    $\ell^+\nu_\ell \ell^{\prime +}\ell^{\prime -}+c.c$ & $b_{\ell\ell}^Wb_{\ell\ell}^Z$ &
    3 & no & $6.34\times 10^3$ & 11.03 \\
    \cline{4-6}
    & 0.0037 && yes & $1.08\times 10^2$ & 3.63 \\
    \hline
    $jj \ell^+\ell^-+c.c$ & $b_{jj}^W b_{\ell\ell}^Z$ & 3 & no & $3.69\times 10^4$ & 26.61 \\
    \cline{4-6}
    & 0.0218 && yes & $6.26\times 10^2$ & 8.75 \\
    \hline
    $\ell^+\nu_\ell jj+c.c$ & $b_{\ell\nu}^W b_{jj}^Z$ & 3 & no & $1.00\times 10^5$ & 43.89 \\
    \cline{4-6}
    & 0.0592 && yes & $1.70\times 10^3$ & 14.43 \\
    \hline
    $jjjj+c.c.$ & $b_{jj}^W b_{jj}^Z$ & 3 & no & $5.84\times 10^5$ & 105.84 \\
    \cline{4-6}
    & 0.3443 && yes & $9.91\times 10^3$ & 34.80 \\
    \hline
    $jjb\bar{b}+c.c.$ & $b_{jj}^W b_{bb}^Z\epsilon_b^2$ & 3 & no & $4.12\times 10^4$ & 28.13 \\
    \cline{4-6}
    & 0.0243 && yes & $7.00\times 10^2$ & 9.25 \\
    \hline
    $\ell^+\nu_\ell b\bar{b}+c.c.$ & $b_{\ell\ell}^W b_{bb}^Z\epsilon_b^2$ & 3 & no & $7.12\times 10^3$ & 11.69 \\
    \cline{4-6}
    & 0.0042 && yes & $1.21\times 10^2$ & 3.84 \\
    \hline\hline
    \multicolumn{6}{|c|}{$pp\rightarrow W^+W^-X$}\\
    \hline
    $\ell_1^+\nu_{\ell_1} \ell_2^{-}\nu_{\ell_2}$ & $(b_{\ell\ell}^W)^2$ &
    3 & no & $3.86\times 10^4$ & 12.73 \\
    \cline{4-6}
    & 0.0121 && yes & $5.14\times 10^2$ & 3.38 \\
    \hline
    $\ell^+\nu_\ell jj$ & $b_{\ell\ell}^W b_{jj}^W$ & 3 & no & $2.24\times 10^5$ & 30.71 \\
    \cline{4-6}
    & 0.0704 && yes & $2.99\times 10^3$ & 8.16 \\
    \hline
    $jjjj$ & $(b_{jj}^W)^2$ & 3 & no & $1.30\times 10^6$ & 74.07 \\
    \cline{4-6}
    & 0.4096 && yes & $1.74\times 10^4$ & 19.68 \\
    \hline
  \end{tabular}
\end{center}
\caption{Cross sections and statistical significance of the model signals
for various decay channels of the analyzed processes. The integrated
luminosity is ${\cal L}=100\;\mbox{fb}^{-1}$.
The reduction factor $r$ is the ratio between the cross sections of the
process with the final state decayed and undecayed.
The cut considered in Table is
$0.7\;\mbox{TeV}\leq m_{34}\leq 1.1\;\mbox{TeV}$.}
\label{tab:AfterDecay}
\end{table}

We can see that the highest yields and statistical significance is in the channels
that involve hadrons. At the same time though, these will be the channels with
the richest and the most complicated backgrounds to deal with. The leptonic
channels exhibit lesser event yield and lesser sensitivity. On the other hand,
they provide cleaner signal and easily detectable particles, like muons.
Thus, when selecting the most appropriate channels for probing our model
it has to be a tradeoff between advantages and disadvantages
of the both worlds. In addition, many of the channels displayed in Table
\ref{tab:AfterDecay} can be combined to reach higher statistics and/or
complementary information on the properties of the model.

\section*{Conclusions}

With the start of the LHC new era in attacking the question of the ESB
mechanism begins.
In this paper we have investigated the potential of the LHC processes
$pp\rightarrow abX$, $ab=t\bar{t},b\bar{b},t\bar{b},W^+Z,W^+W^-$,
to probe a new vector $SU(2)_V$ triplet which might exist as a part
of new physics responsible for ESB and, as far as fermions are concerned,
it interacts directly to the third quark generation only.
It is possible that new physics behind ESB
can be effectively described by the HLS approach. Then the mixing of
the vector resonances with the electroweak gauge bosons induces
also interactions of $\rho$ to lighter SM fermions.

The processes in which $\rho$ interacts only
to quarks of the third generation
should be more sensitive to new physics than
those which run through $\rho$-to-light-fermion vertices.
However, at the LHC the former processes are often overwhelmed by a huge
gluon-gluon background. We have seen this in the case of
$pp\rightarrow t\bar{t}X$ and $pp\rightarrow b\bar{b}X$.
We have demonstrated that also the processes where $\rho$ couples
to the light quarks in protons have a significant potential to
distinguish our model from the light Higgs SM.
We have calculated the statistical significance of the model signal
for various decay channels of the final states. However, to
decide whether and which
of these channels can recognize the model at the LHC
the further analysis, which would
include the study of backgrounds and the detector reconstruction efficiency,
is necessary.
This is out of the scope of this paper, though. The results in this paper
should help decide which of the processes are the best candidates for such further
analyses.


\end{document}